\documentclass[fleqn,12pt,twoside]{article}
\usepackage[headings]{espcrc1}

\readRCS
$Id: espcrc1.tex,v 1.2 2004/02/24 11:22:11 spepping Exp $
\usepackage{graphicx}

\newcommand{\AmS}{{\protect\the\textfont2
  A\kern-.1667em\lower.5ex\hbox{M}\kern-.125emS}}

\title{3D Relativistic Hydrodynamic Computations Using 
 Lattice-QCD-Inspired Equations of State}

\author{Yogiro~Hama\address[USP]{Instituto de F\'{\i}sica, 
                Universidade de S\~ao Paulo, Brazil},  
        Rone~P.G.~Andrade\addressmark,
        Fr\'ed\'erique~Grassi\addressmark[USP],  
        Ot\'avio~Socolowski~Jr.\address{Instituto Tecnol\'ogico 
                da Aeron\'autica, Brazil},
        Takeshi~Kodama\address[UFRJ]{Instituto de F\'{\i}sica, 
                Universidade Federal do Rio de Janeiro, 
                Brazil}, 
        Bernardo~Tavares\addressmark  
        and     
        Sandra~S.~Padula\address{Instituto de F\'{\i}sica 
        Te\'orica, Universidade Estadual Paulista, Brazil}}
       
\runtitle{3D Relativistic Hydrodynamic Computations ...}
\runauthor{Y. Hama {\it et al.}}

\begin{document}

\maketitle 

\begin{abstract}
In this communication, we report results of three-dimensional 
hydrodynamic computations, by using equations of state with a 
critical end point as suggested by the lattice QCD. Some of the 
results are an increase of the multiplicity in the mid-rapidity 
region and a larger elliptic-flow parameter $v_2\,$. 
We discuss also the effcts of the initial-condition 
fluctuations and the continuous emission. 
\end{abstract}

\section{INTRODUCTION}

Nowadays, it is widely accepted that hydrodynamics is a 
successful approach for describing the bulk of the collective 
flow in high-energy nuclear collisions \cite{BMuller}. 
The basic assumption in hydrodynamical models is the local 
thermal equilibrium. Once this condition is satisfied, all the 
thermodynamical relations should be valid in each space-time 
point. The properties of the matter formed in high-energy 
collisions are then specified by some equations of state (EoS). 
Thus, one of the main objects of hydrodynamical approach is to 
determine which are the EoS that consistently reproduce the 
observed quantities. 

In high-energy nucleus-nucleus collisions, one often uses  EoS 
with a first-order phase transition, connecting a 
high-temperature QGP phase of the system with a low-temperature 
hadronic phase. However, lattice QCD studies showed that the 
transition line between QGP and hadron phase has a critical end 
point and for small net baryon surplus the transition is of 
crossover type~\cite{LQCD}. 
So, we would like to learn what are the consequences of these  
results on the hydrodynamics and on the observable quantities. 

We shall begin showing, in the next Section, how the critical end 
point could be implemented for the sake of phenomenological 
computations. 
Besides the EoS, the ingredients of any hydrodynamic approach are 
the equations of motion, the initial conditions and some 
decoupling prescription. In Sec. \ref{ingredients}, we shall 
discuss how the initial conditions are chosen in our studies; 
then, how we solve the hydrodynamic equations; and finally 
which is the decoupling prescription. In Sec. \ref{results}, 
we shall show some of the results of our studies. 
Finally, summaries of conclusions and outlook are given in 
Sec. \ref{conclusions}. 

\section{PARAMETRIZATION OF EQUATIONS OF STATE} 
\label{EoS}

As mentioned above, one often introduces EoS with a first-order 
phase transition, connecting a QGP phase of the system, usually 
described by the MIT bag model, with a hadronic phase, depicted 
as a resonance gas. A detailed account of such EoS may be 
found, for instance, in \cite{review}. 
We start from these EoS, in order to get a phenomenological parametrization of those suggested by lattice QCD. 
 
\begin{figure}[h]
\vspace*{-.5cm} 
\begin{minipage}[h]{75mm}
\includegraphics*[scale=.31]{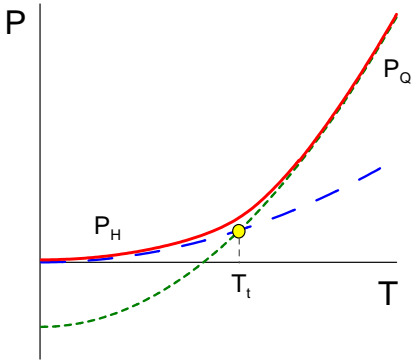} 
\vspace*{-.3cm} 
\caption{Illustration of Eqs.(\ref{pressure}) and (\ref{delta}).}
\label{fig:ptCP2}
\end{minipage} 
\hspace{\fill}
\begin{minipage}[t]{75mm} 
\vspace*{-3.5cm} 
could choose some function which becomes exactly 0 for 
$\mu_b>\mu_c$ to guarantee the first-order phase transition 
there, but for practical purpose this is not necessary. As 
will be shown below, our choice represented by Eq.(\ref{delta}) 
is enough. We shall designate the equations of state 
given above, with $\delta_0\not=0$, CP EoS. 

\quad Let us compare in Figure \ref{fig:EoS1} below, the 
temperature dependences of the energy and entropy densities,  $\varepsilon$ and $s$, and the pressure $P$, given by the two 
sets of EoS defined above. One can see that the cross-\hfilneg\ 

\end{minipage} 
\end{figure}

Let us denote by $P_Q$ the pressure given by the MIT bag model 
and $P_H$ the one corresponding to the hadronic resonance gas. 
Given a value $\mu_b$ of the baryonic chemical potential, we 
write for the pressure $P$ the equation (see Figure 
\ref{fig:ptCP2}) 
\begin{equation}
 (P-P_Q)(P-P_H)={\delta(\mu_b)}\,, \label{pressure}
\end{equation} 
where 
\begin{equation}
\delta(\mu_b)=\delta_0\exp\left[-(\mu_b/\mu_c)^2\right],
\quad\quad\hbox{with} 
\quad\quad\mu_c=\hbox{critical chemical potential}. 
\label{delta} 
\end{equation} 

By solving the equation above and using thermodynamical 
relations, we obtain 
\begin{eqnarray}
  P  &=& \lambda P_H+(1-\lambda)P_Q 
        +\frac{2\delta}{\sqrt{(P_Q-P_H)^2+4\delta}}\ ,\\
  s  &=& \lambda s_H+(1-\lambda) s_Q\,, \\
  n_b&=& \lambda n_H+(1-\lambda) n_Q 
        -\frac{2\,(\mu/\mu_c^2)\,\delta}
         {\sqrt{(P_Q-P_H)^2+4\delta}}\ , \\
  \epsilon &=& \lambda\epsilon_H+(1-\lambda)\epsilon_Q 
        -\frac{2\,\left[1+(\mu/\mu_c)^2\right]\,\delta}
         {\sqrt{(P_Q-P_H)^2+4\delta}}\ , \\
  \hbox{where}\quad \lambda&\equiv&{1\over2}
     \left[1-(P_Q-P_H)/\sqrt{(P_Q-P_H)^2+4\delta}\right]\ . \end{eqnarray}
Observe that if $\delta_0=0$, we recover the EoS with the 
first-order phase transition described above. In the discussion 
below, we call them 1OPT EoS. 
As seen in Figure \ref{fig:ptCP2}, when $\delta(\mu_b)\not=0$, 
the transition from hadron phase to QGP is smooth. In the 
right-hand side of Eq.(\ref{pressure}), we\hfilneg\ 

\begin{figure}[!t]
\vspace*{-1.5cm}
\includegraphics[height=13.5cm,width=20.5cm]{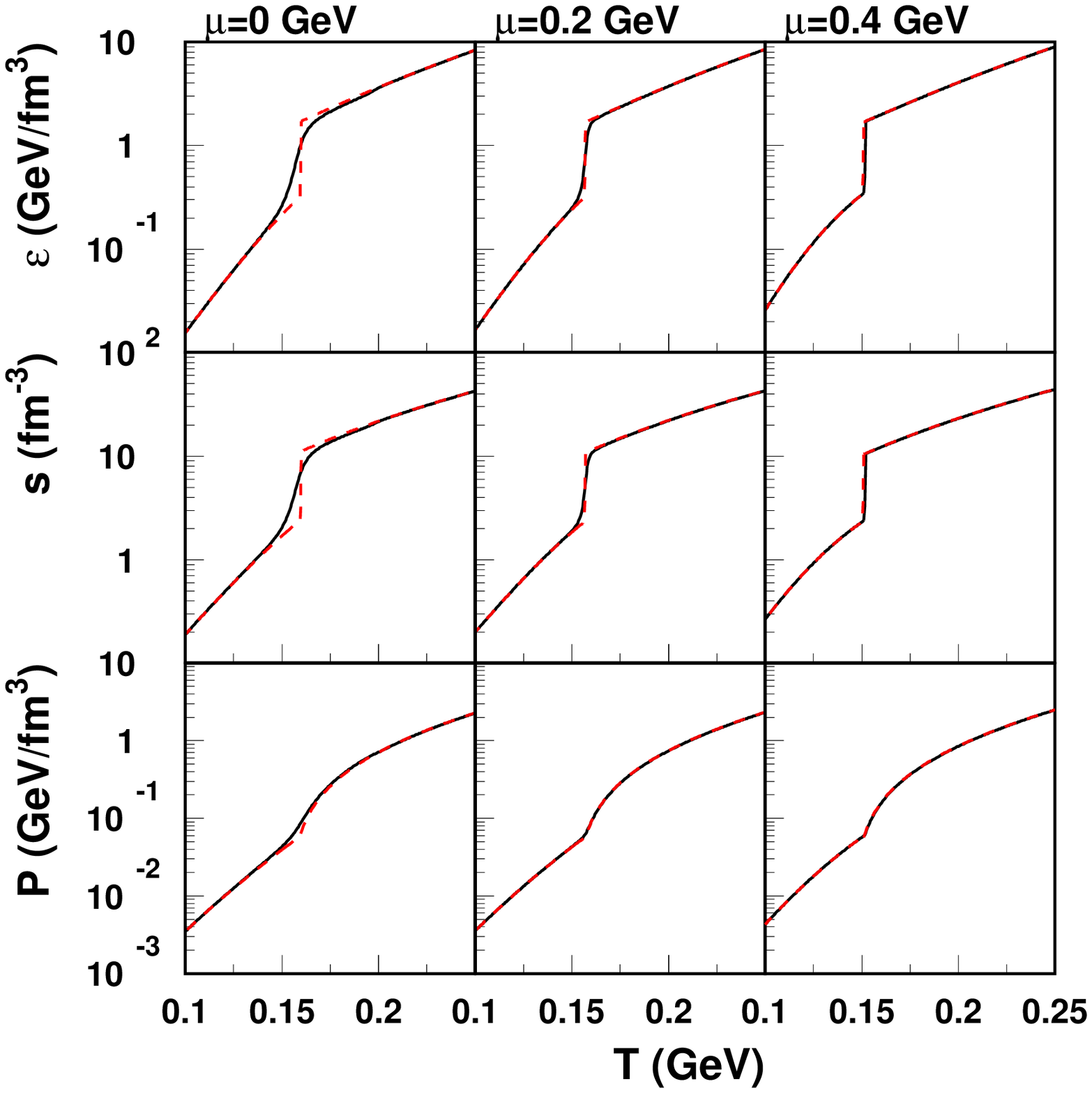} 
\vspace{-2.cm} 
\caption{A comparison of $\varepsilon(T)$, $s(T)$ and $P(T)$ 
 as given by our parametrization with a critical point (solid 
 lines) and those with a first-order phase transition (dashed 
 lines).}
\label{fig:EoS1}
\vspace{-.5cm} 
\vspace{-.3cm}
\includegraphics[height=10.cm,width=20.5cm]{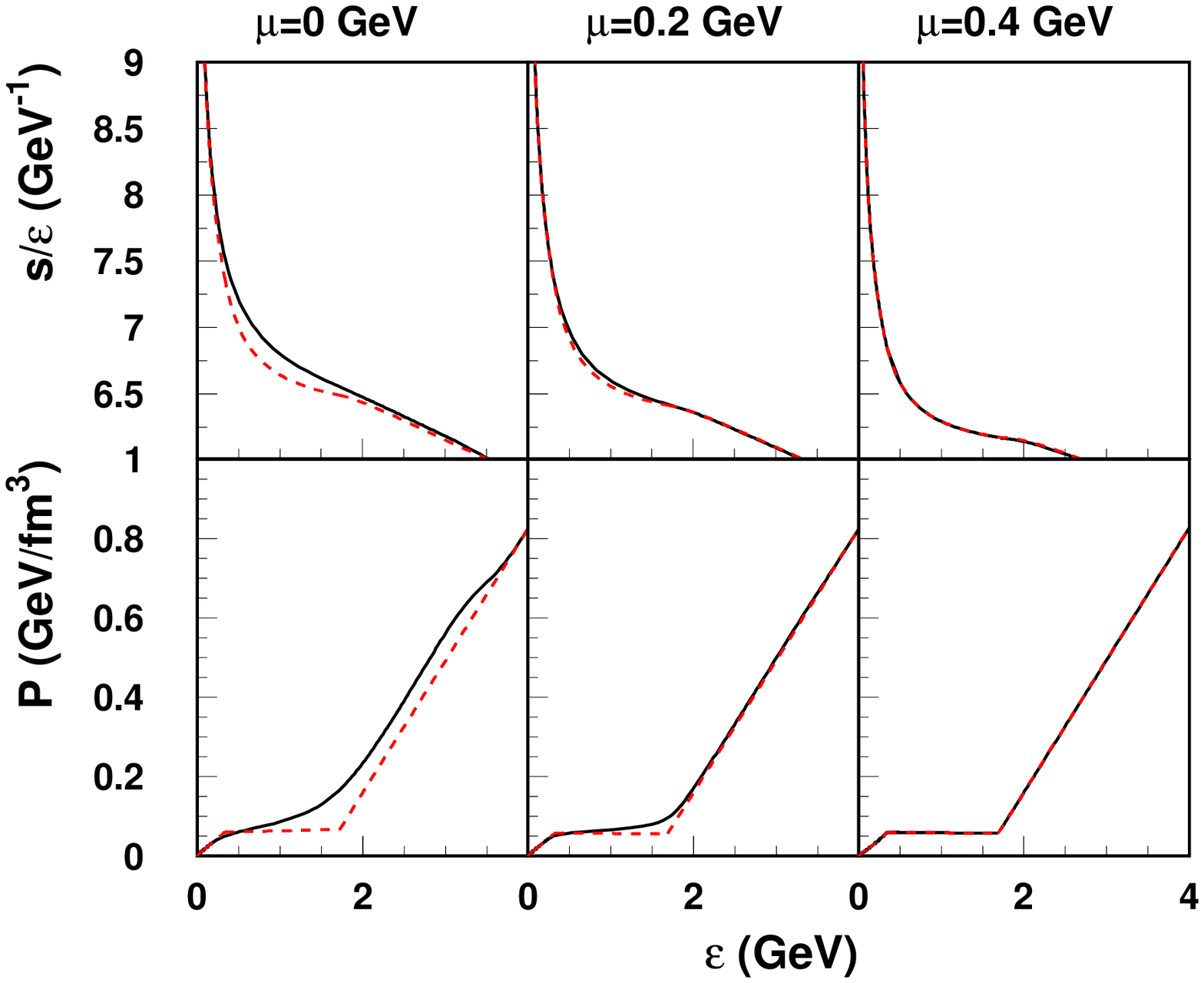}
\vspace{-1.5cm} 
\caption{Plots of $s/\varepsilon$ and $P$ as function of 
 $\varepsilon$ for the two EoS shown in Figure~\ref{fig:EoS1}.}
\label{fig:EoS2}
\end{figure}
\newpage

\noindent over behavior is correctly reproduced by our 
parametrization for CP EoS, while finite jumps in 
$\varepsilon(T)$ and $s(T)$ are exhibited by 1OPT EoS, when $T$ 
crosses the transiton temperature. It is also seen, as 
mentioned above, that at $\mu_b\sim0.4\,$GeV the two EoS are 
indistinguishable. 

Now, since in a real collision what is directly given is the 
energy distribution at a certain initial time (besides 
baryon number distribution, charge distribution, strangeness 
distribution, {\it etc.}), whereas the temperature is defined 
with the use of the former, it would be nice to compare the 
two sets of EoS, by plotting several quantities as function of 
$\varepsilon$. We do this in Figure \ref{fig:EoS2}. One 
immediately sees there some remarkable differences between the 
two sets of EoS: $i\,$) naturally the pressure is not 
constant for CP EoS in the crossover region; $ii\,$) moreover, 
the entropy is larger. We will see in Sec.4 that these 
characteritics affect the observed quantities in non-negligible 
way.  

\section{OTHER INGREDIENTS OF OUR HYDRODYNAMIC MODEL}
\label{ingredients}

Besides the equations of state, the other ingredients of a 
hydrodynamic model are the initial conditions, the equations 
of motion and some decoupling prescription. Here we shall 
discuss how these elements are chosen in our studies. 

\subsection{Initial Conditions} 

In usual hydrodynamic approach, one assumes some highly 
symmetric and smooth initial conditions (IC). However, since 
our systems are small, large event-by-event fluctuations are 
expected in real collisions, so this effect should be taken 
into account. Remark that this might happen even if the impact 
parameter could be maintained fixed. 

\begin{figure}[b] 
\vspace*{-1.cm}
\begin{minipage}[h]{10mm}
\end{minipage}
\hspace{\fill}
\begin{minipage}[h]{75mm}
\vspace*{.cm}
    \includegraphics[angle=270,width=10cm]{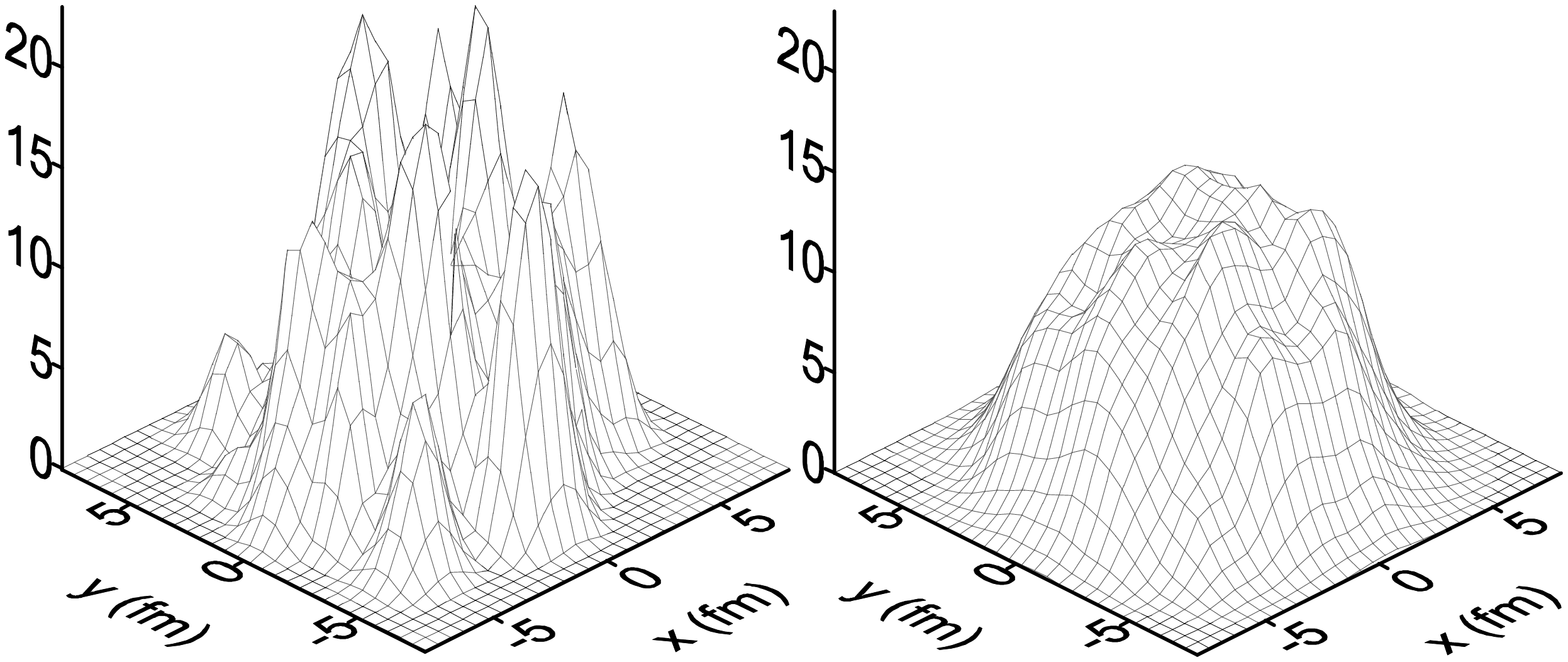}
\end{minipage}    
\hspace{\fill}
\begin{minipage}[h]{20mm}
\end{minipage}
\vspace*{-.4cm}
\caption{The energy density is plotted in units of GeV/fm$^3$ 
 at the initial instant. 
 Left: one random event. Right: average over 30 random events 
 (corresponding to the smooth initial conditions in the usual 
 hydro approach).}
\label{fig:IC}
\end{figure}

Many simulators, based on microscopic models, {\it e.g.} 
HIJING \cite{hijing}, VNI \cite{vni}, URASiMA \cite{urasima}, 
NeXuS \cite{nexus}, $\cdots$, show such event-by-event 
fluctuations. As an example we show in Figure \ref{fig:IC} the 
energy density for central Au+Au collisions at 130A GeV, given 
by NeXuS simulator \cite{nexus}, at mid-rapidity. In our 
approach, we use both fluctuating and averaged IC. Some 
consequences of such fluctuations have been discussed 
elsewhere. We shall discuss some others in Sec.\ref{results}. 

\subsection{Equations of Motion} 

The equations of motion of hydrodynamics are the continuity 
equations expressing the energy-momentum conservation, the 
baryon-number conservation, and other conservation laws, 
corresponding to several charges. Here, for the  sake of 
simplicity, we shall consider only the energy-momentum and the 
baryon number. Since our IC are entirely arbitrary, without any 
symmetry, as discussed above, the only way to solve the 
equations is through numerical computations. We have developed 
a numerical code called SPheRIO ({\bf S}moothed {\bf P}article 
{\bf h}ydrodynamic {\bf e}volution of {\bf R}elativistic heavy 
{\bf IO}n collisions) \cite{spherio}, based on the so called 
Smoothed-Paricle Hydrodynamics (SPH) algorithm \cite{sph}. 

\subsection{Decoupling Prescription} 
\label{ce} 
In hydrodynamic treatment of high-energy nuclear collisions, 
one often assumes decoupling on a sharply defined hypersurface, 
usually characterized by a constant temperature $T_{fo}$. We 
call this {\it Sudden Freeze Out} (FO). However, our systems 
are small, so particles may escape from a layer with thickness 
comparable with the systems sizes. We have proposed an 
alternative decoupling prescription that we call 
{\it Continuous Emission} (CE)~\cite{ce} which, as compared to the usual sudden freeze out, we believe closer to what happens 
in the actual collisions. We introduce, at each space-time 
point $x^\mu$, a certain momentum-dependent escaping 
probability 
\begin{equation} 
 {\cal P}(x,k)
   =\exp\left[-\int_\tau^\infty \rho(x^\prime)\,\sigma v\;
    \mathrm{d}\tau^\prime\right].  
\label{eq:prob1} 
\end{equation} 

To implement this prescription in our SPheRIO code, we had to 
introduce some approximation to make the computation 
practicable. First, we take ${\cal P}$ on the average, 
{\it i.e.}, 
\begin{equation} 
  {\cal P}(x,k)\rightarrow\langle{\cal P}(x,k)\rangle 
     \equiv {\cal P}(x) 
\end{equation} 
and then we approximate linearly the density 
$\rho(x^\prime)=\alpha\, s(x^\prime)$
in Eq.(\ref{eq:prob1}). Thus, 
\begin{equation} 
 {\cal P}(x,k)\rightarrow{\cal P}(x)
     =\exp\left(-\kappa     
      \,\frac{s^2}{|\mathrm{d}s/\mathrm{d}\tau|}\right), 
\label{eq:prob} 
\end{equation} 
where $\kappa = 0.5\,\alpha\,\langle\sigma v\rangle$ is 
estimated to be 0.3$\,$, corresponding to 
$\langle\sigma v\rangle\approx$ 2~fm$^2$. 

\begin{figure}[!h]
\vspace*{-3.2cm} 
\begin{minipage}[h]{77mm}
\includegraphics*[scale=0.38]{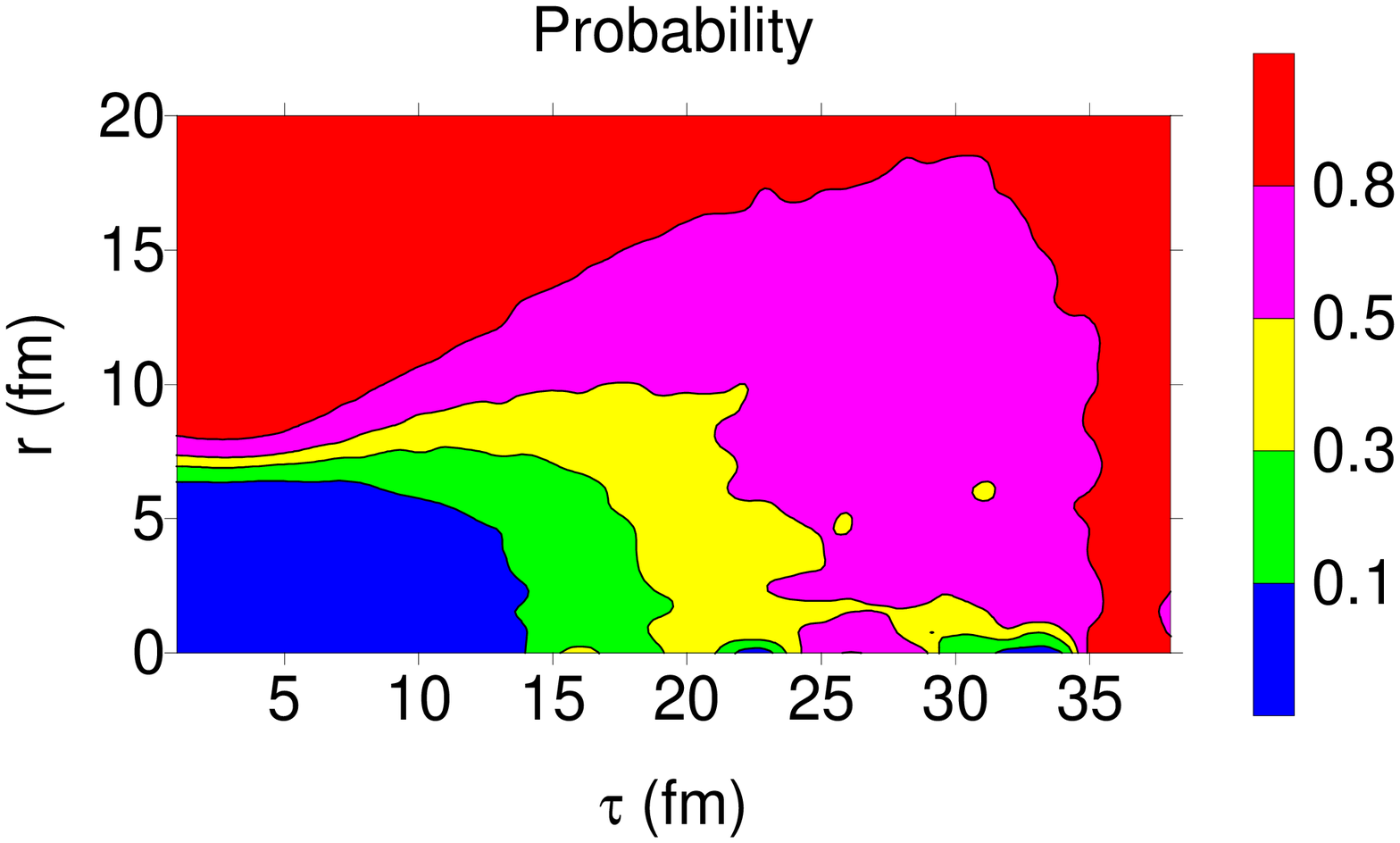}
\vspace*{-4.4cm} 
\caption{Time evolution of the probability as given by 
 Eq.(\ref{eq:prob}) for the most central Au+Au collisions at 
 130$\,$A GeV, in the mid-rapidity plane for averaged IC.}
\label{fig:prob}
\end{minipage}
\hspace{\fill}
\begin{minipage}[h]{77mm}
\vspace*{-1.cm} 
\includegraphics*[scale=0.38]{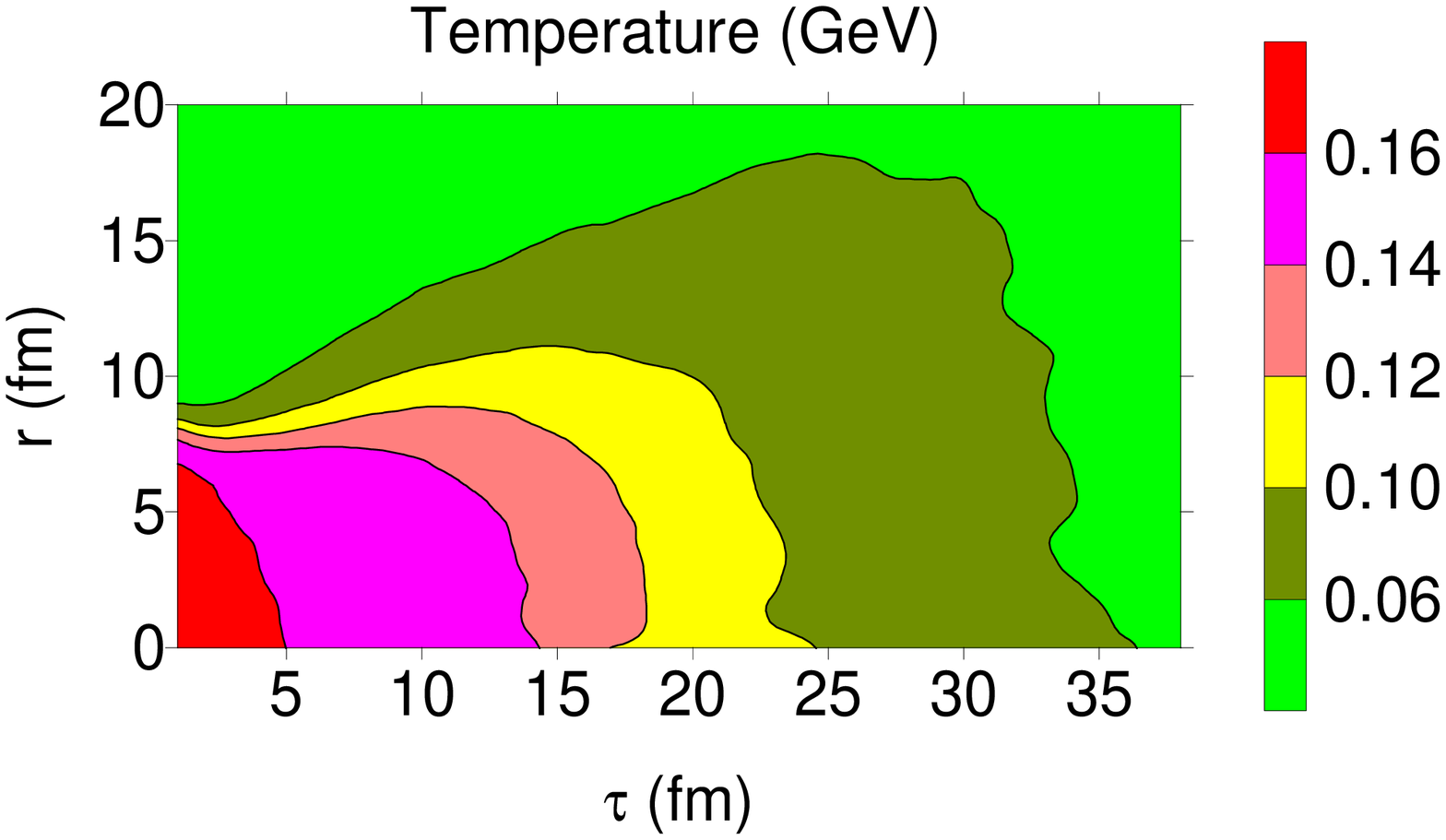}
\vspace*{-4.4cm} 
\caption{Corresponding time evolution of the temperature.} 
\label{fig:temperature} 
\end{minipage}
\vspace*{-.6cm} 
\end{figure}

We show, in Figure \ref{fig:prob}, the time ($\tau$) evolution 
of the probability ${\cal P}$, estimated by this expression, 
in the mid-rapidity plane for the most central Au+Au collisions 
at 130$\,$A GeV. For comparison, we show, in Figure 
\ref{fig:temperature}, the corresponding time evolution of the 
temperature $T$. One sees that, whereas in {\it Sudden Freeze 
Out}, particles are emitted from a constant $T$ line of 
Figure \ref{fig:temperature}, in {\it Continuous Emission}, 
they are emitted according to the probability ${\cal P}$, so 
from a difuse space region and during a larger time interval. 
It will be shown in Sec. \ref{results} that this difference 
gives important changes in some observables.  
\eject   
 
\section{RESULTS}
\label{results} 

Let us now show results of computation of some observables, as 
described above, for Au+Au collisions at 200A GeV. We start  
computing the pseudo-rapidity and the transverse-momentum  
distributions for charged particles, to fix the parameters. 
Then, the elliptic-flow parameter $v_2$ and HBT radii are 
computed in fit-parameter free way. 

\subsection{Pseudo-rapidity distribution} 

\begin{figure}[!hb]
\begin{minipage}[h]{80mm}
\vspace*{-1.4cm}
\includegraphics*[scale=0.33]{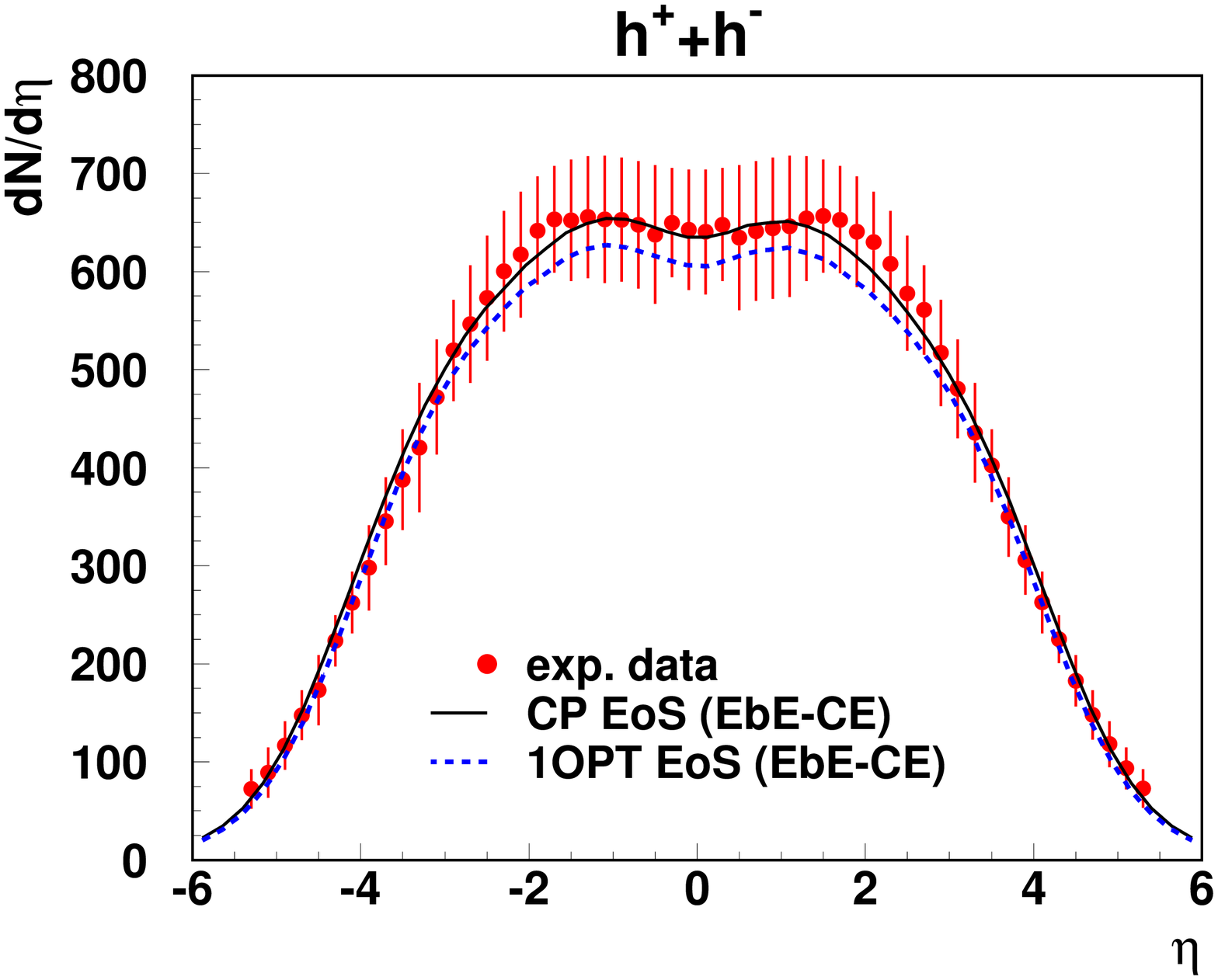}
\vspace*{-1.5cm} 
\caption{$\eta$ distributions for the most central Au+Au at 
 200A GeV. A comparison of CP EoS (solid line) vs. 1OPT EoS 
 (dashed line). The data are from PHOBOS Collab.\cite{phobos1}.}
\label{fig:eta1}
\end{minipage}
\hspace{\fill}
\begin{minipage}[h]{75mm}
\vspace*{-1.4cm}
\includegraphics*[scale=0.33]{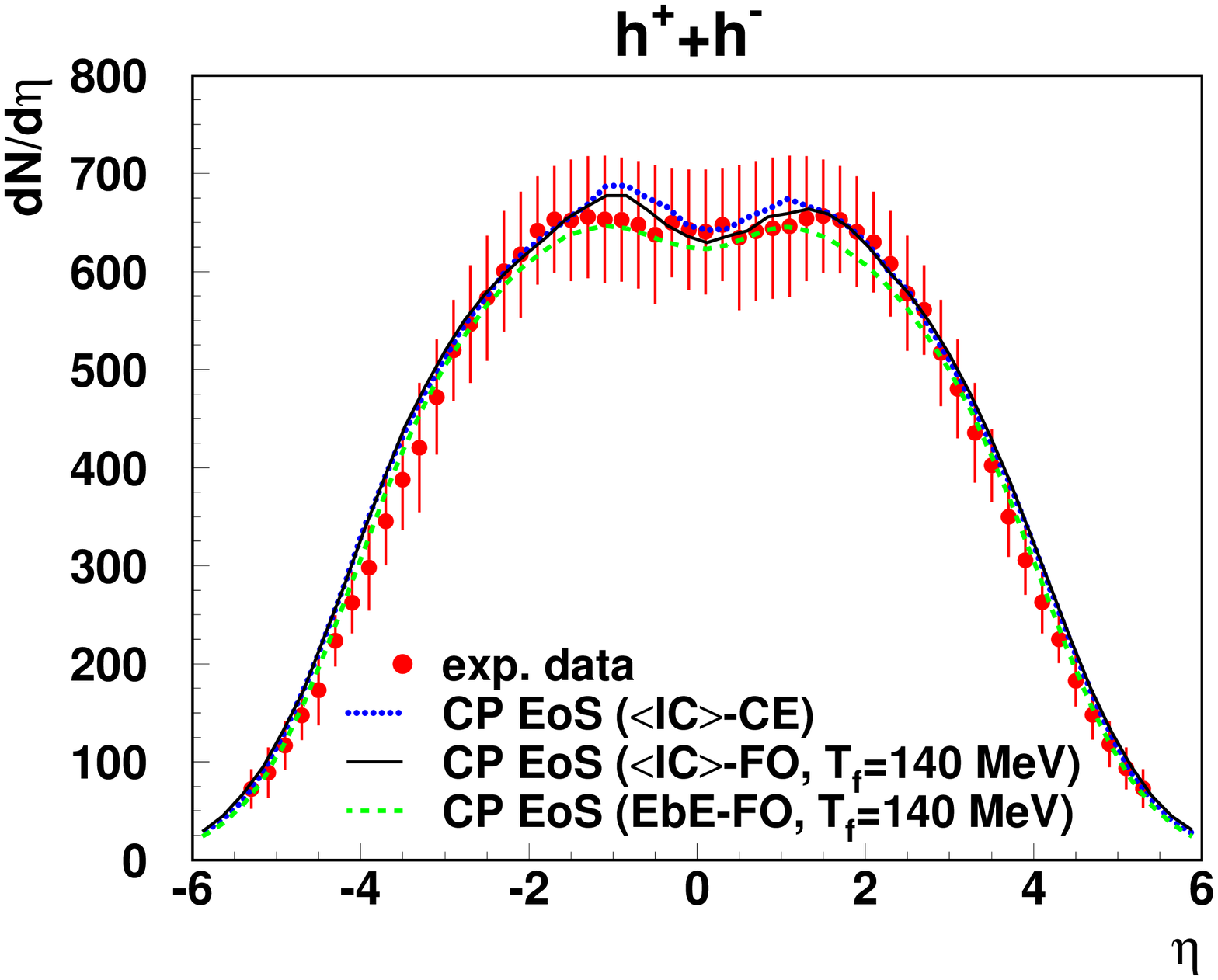}
\vspace*{-1.5cm} 
\caption{$\eta$ distributions for the most central Au+Au at 
 200A GeV, computed with CP EoS, with three different 
 combinations of IC and decoupling prescriptions.}
\label{fig:eta}
\end{minipage}
\end{figure}

In Sec. \ref{EoS}, we showed that the inclusion of a critical 
end poit in the first-order phase-transition line increases 
the entropy per energy, as compared with 1OPT EoS. This means 
that, given the same total energy, the multiplicity is larger 
for CP EoS case than for 1OPT EoS one. Figure \ref{fig:eta1} 
shows clearly that this happens, especially in the mid-rapidity 
region. 

Now, once the equations of state are chosen, what is the 
effect of the fluctuating initial conditions for the same 
decoupling prescription? This effect has already been 
discussed\hfilneg\  

\noindent earlier~\cite{review} and shown in Figure 
\ref{fig:eta}, namely, the multiplicity is smaller if the IC 
fluctuation is taken into account and average computed after 
the decoupling, as compared with the result obtained with 
smooth averaged IC. The curve with continuous emission is also 
shown there, reproducing equally well the data. Here, the 
parameter $\kappa$ has been fixed as explained in 
Subsec.~\ref{ce}. 

\subsection{Transverse-Momentum Distribution} 

\begin{figure}[!hb]
\begin{minipage}[h]{80mm}
\vspace*{-1.3cm}
\includegraphics*[scale=0.34]{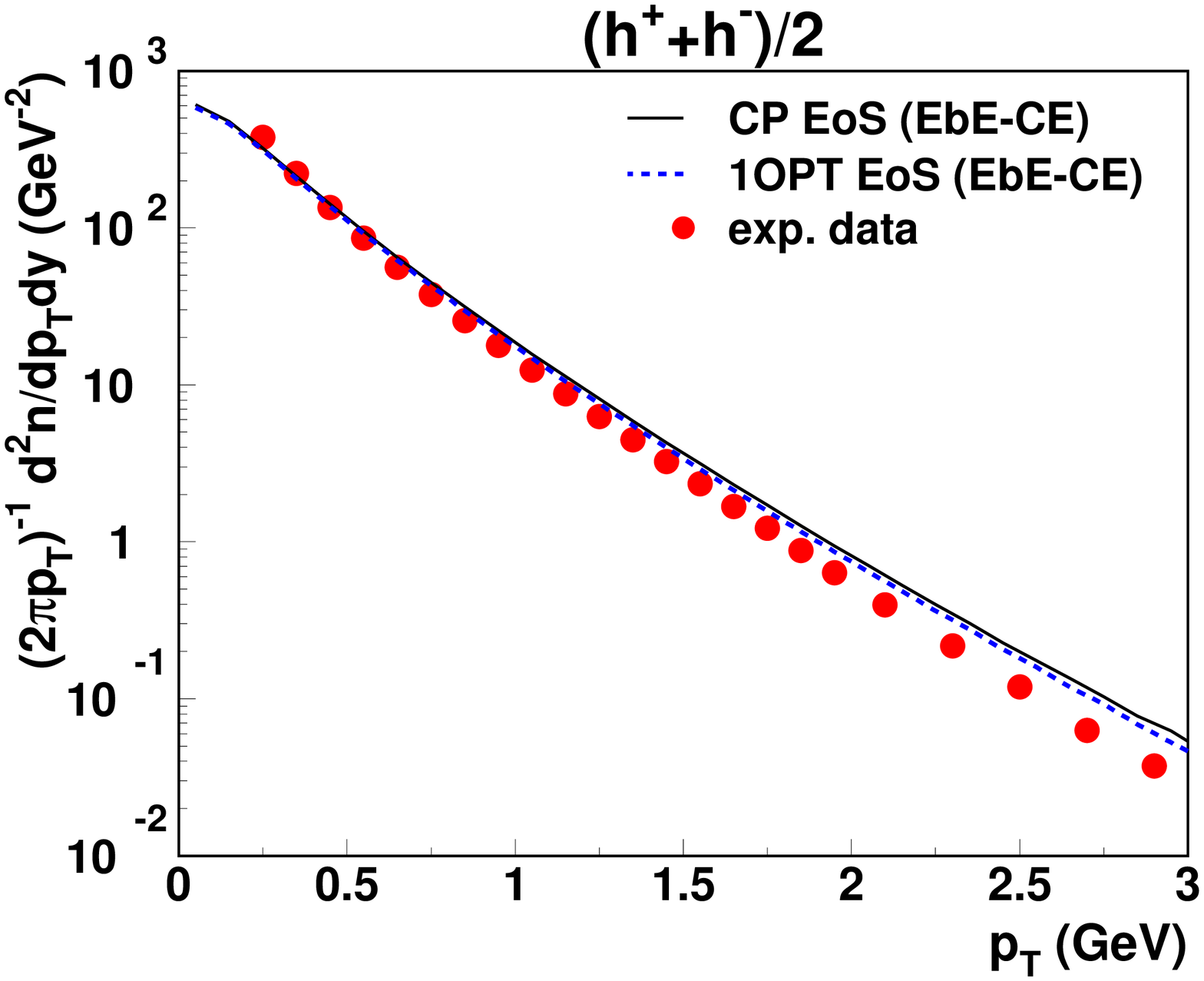}
\vspace*{-1.5cm} 
\caption{$p_T$ distributions for the most central Au+Au at 
 200A GeV. A comparison of CP EoS (solid line) vs. 1OPT EoS 
 (dashed line). The data are from PHOBOS Collab.\cite{phobos2}.}
\label{fig:pT1}
\end{minipage}
\hspace{\fill}
\begin{minipage}[h]{75mm}
\vspace*{-1.3cm}
\includegraphics*[scale=0.34]{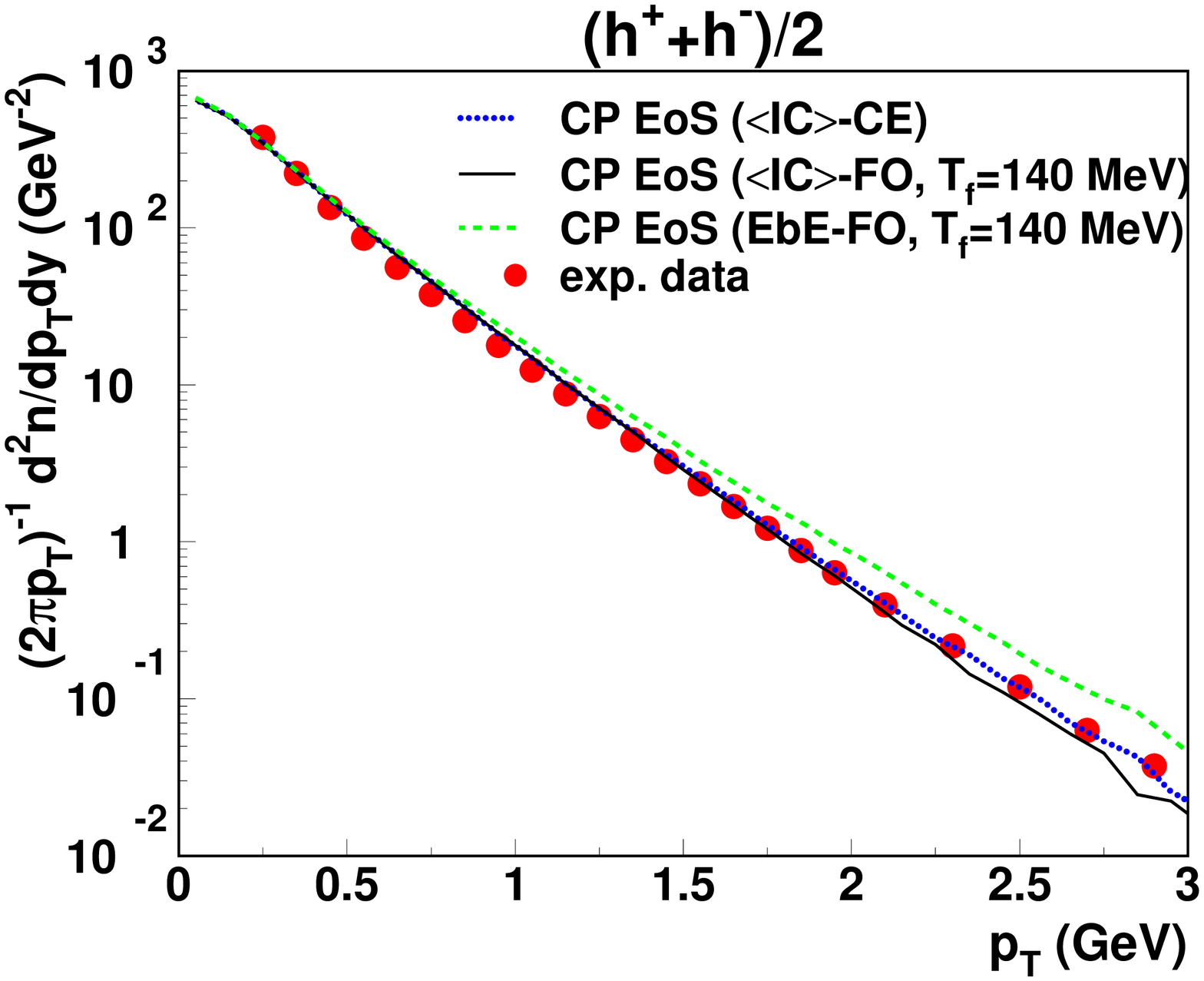}
\vspace*{-1.5cm} 
\caption{$p_T$ distributions for the most central Au+Au at 
 200A GeV, computed with CP EoS, with three different 
 combinations of IC and decoupling prescriptions.}
\label{fig:pT}
\end{minipage}
\vspace*{-.5cm} 
\end{figure}

As discussed in Sec. \ref{EoS}, since the pressure does not 
remain constant in the crossover region, we expect that the 
transverse acceleration is larger for CP EoS, as compared with 
1OPT EoS case. In effect, Figure \ref{fig:pT1} does show that 
$p_T$ distribution is flatter for CP EoS, but the difference 
is small. 

We show, in Figure \ref{fig:pT}, three different combinations 
of IC and decoupling prescriptions, corresponding to CP EoS.  
The curve with the event-by-event fluctuating IC is flatter 
than the one corresponding to the averaged IC, both with sudden 
freezeout, probably because the initial expansion in the former 
is more violent, due to the bumpy structure with 
high-density blobs, as seen in Figure \ref{fig:IC} in this 
case. The freezeout temperature suggested by $\eta$ and $p_T$ 
distributions turned out to be $T_f\simeq135-140\,$MeV.  

\subsection{Elliptic-Flow Parameter $v_2$} 

\begin{figure}[!h]
\begin{minipage}[h]{80mm}
\includegraphics*[scale=0.325]{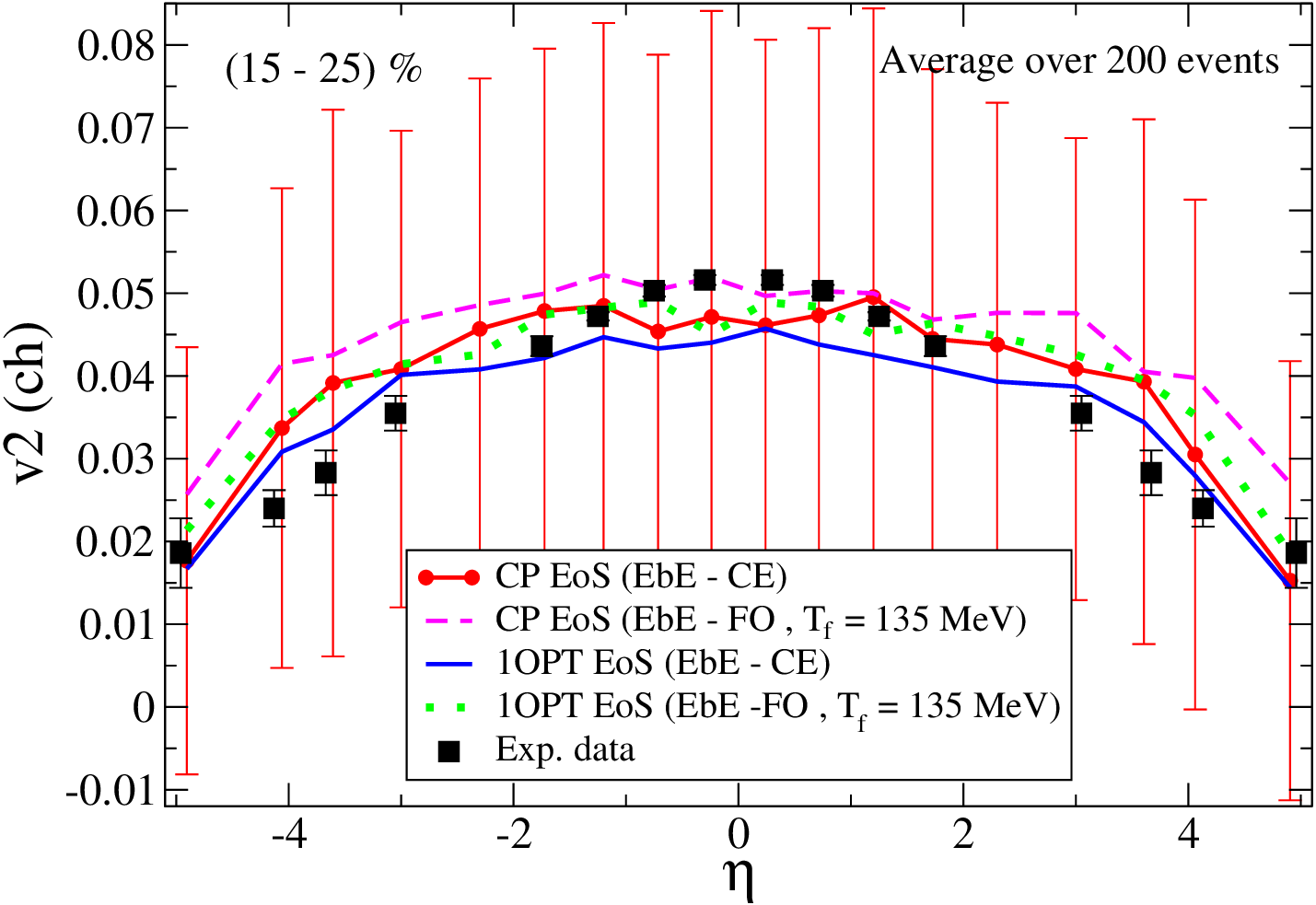} 
\vspace*{-.5cm}
\caption{Pseudo-rapidity distribution of $v_2$ for charged 
 particles in the centrality $(15-25)$\% Au+Au at 200A GeV, 
 computed with event-by-event fluctuating IC. The vertical bars 
 indicate dispersions. The data are from PHOBOS 
 Collab.\cite{phobos3}.}
\label{fig:v2}
\end{minipage}
\hspace{\fill}
\begin{minipage}[h]{75mm}
\vspace*{-1.2cm}
\includegraphics*[scale=0.315]{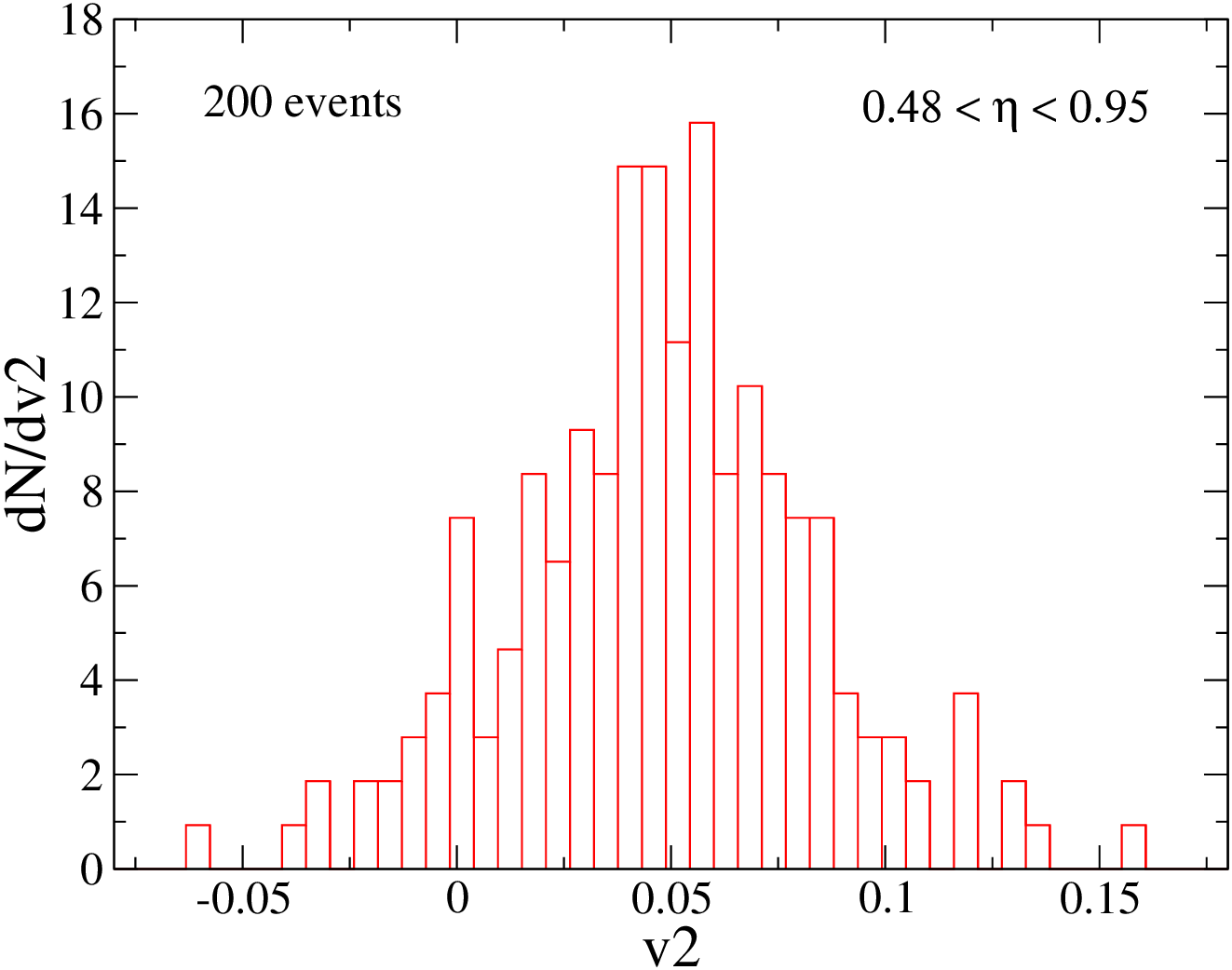} 
\vspace*{-.5cm}
\caption{Event-by-event $v_2$ distribution in the interval 
 $0.48<\eta<0.95$ and corresponding to the same events as in 
 Figure \ref{fig:v2}.}
\label{fig:dNdv2}
\end{minipage}
\end{figure}

Here, we show our results for the pseudo-rapidity distribution 
of the elliptic-flow parameter $v_2$ for Au+Au collisions at 
200A GeV. As seen in Figure \ref{fig:v2}, CP EoS gives larger 
$v_2\,$, as a consequence of larger acceleration in this case 
as discussed in Sec.\ref{EoS}. Notice that the continuous 
emission makes the curves narrower, as a consequence of earlier 
emission of particles, so smaller acceleration, at 
large-$\vert\eta\vert$ regions. See more detils in\cite{rone}. 
When the IC fluctuations are taken into account, the resulting 
fluctuations of $v_2$ become large, as seen in Figures 
\ref{fig:v2} and \ref{fig:dNdv2}. It would be nice to measure 
such a $v_2$ distribution, which would discriminate among 
several microscopic models for the initial stage of nuclear 
collisions. 

\subsection{HBT Radii} 

\begin{figure}[!h]
\begin{minipage}[h]{80mm}
\includegraphics*[scale=0.38]{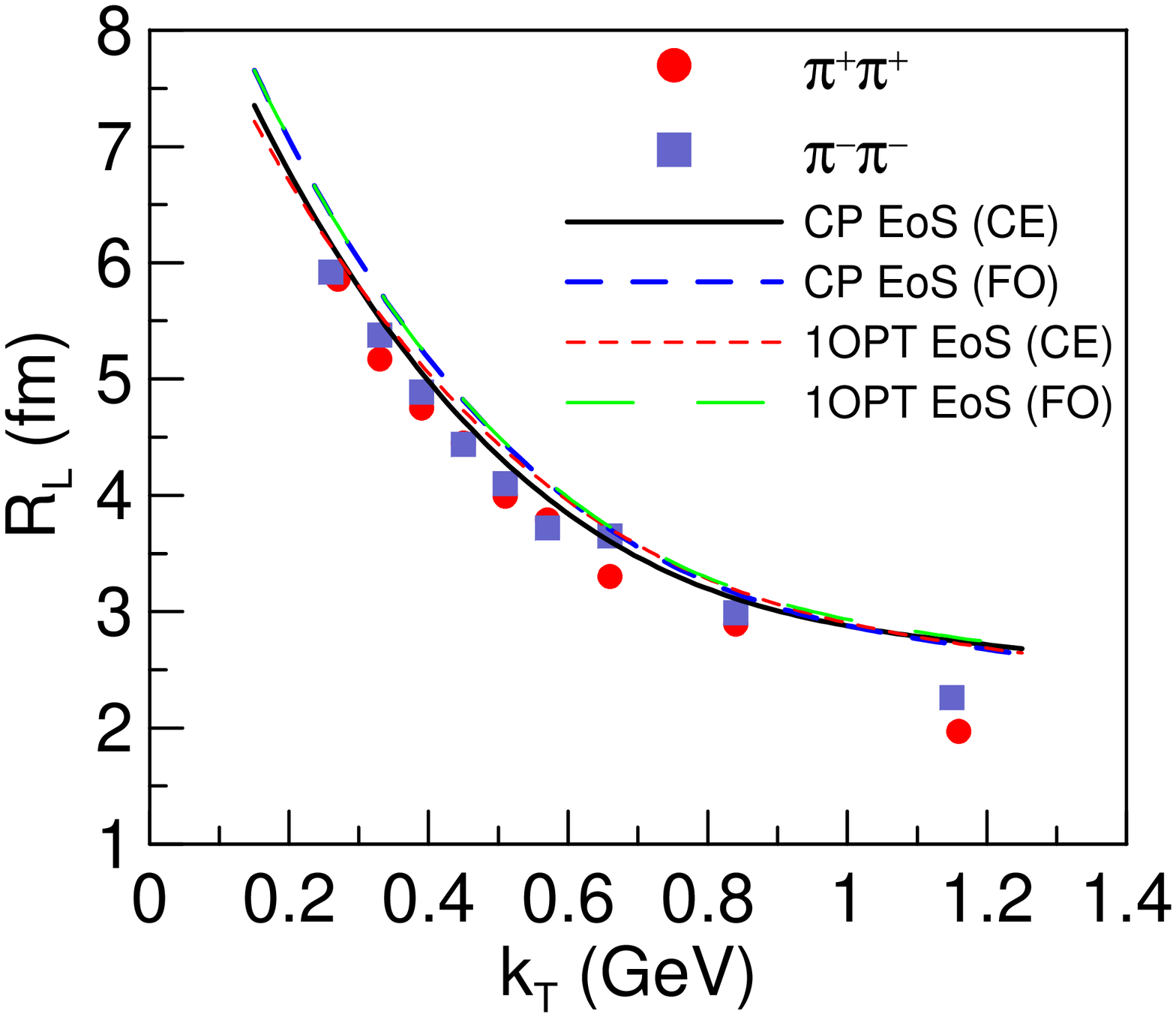}
\vspace*{-.5cm}
\caption{$k_T$ depenence of HBT radius $R_L$ for pions in 
 the most central Au+Au at 200A GeV, computed with 
 event-by-event fluctuating IC. The data are from PHENIX 
 Collab.\cite{phenix}.}
\label{fig:RL}
\end{minipage}
\hspace{\fill}
\begin{minipage}[h]{75mm}
\vspace*{.cm}
\includegraphics*[scale=0.38]{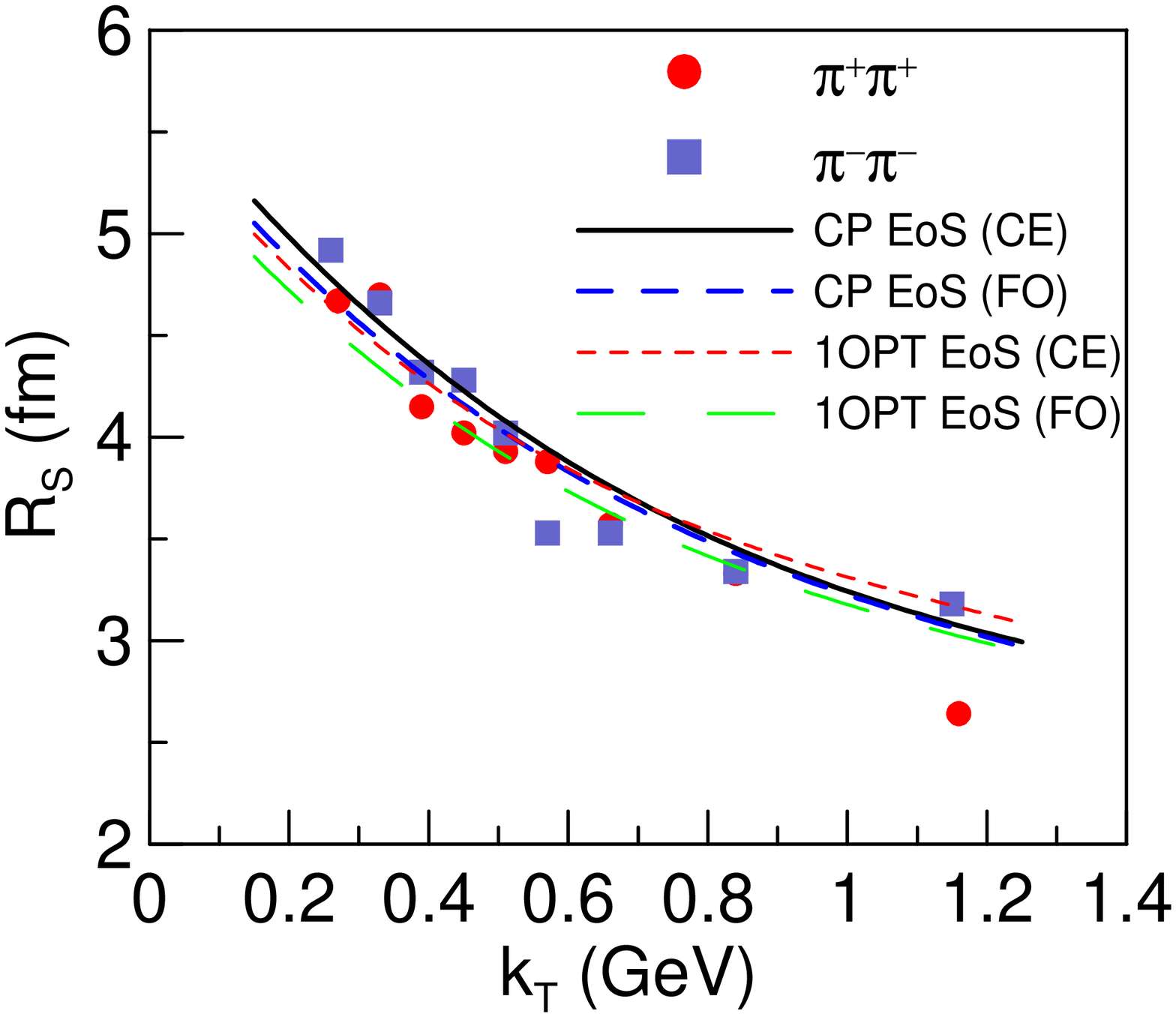}
\vspace*{-1.cm}
\caption{$k_T$ depenence of HBT radius $R_s$ for pions in 
 the most central Au+Au at 200A GeV, computed with 
 event-by-event fluctuating IC. The data are from PHENIX 
 Collab.\cite{phenix}.}
\label{fig:Rs}
\end{minipage}
\end{figure}

\begin{figure}[!h]
\begin{minipage}[h]{80mm}
\vspace*{.cm}
\includegraphics*[scale=0.38]{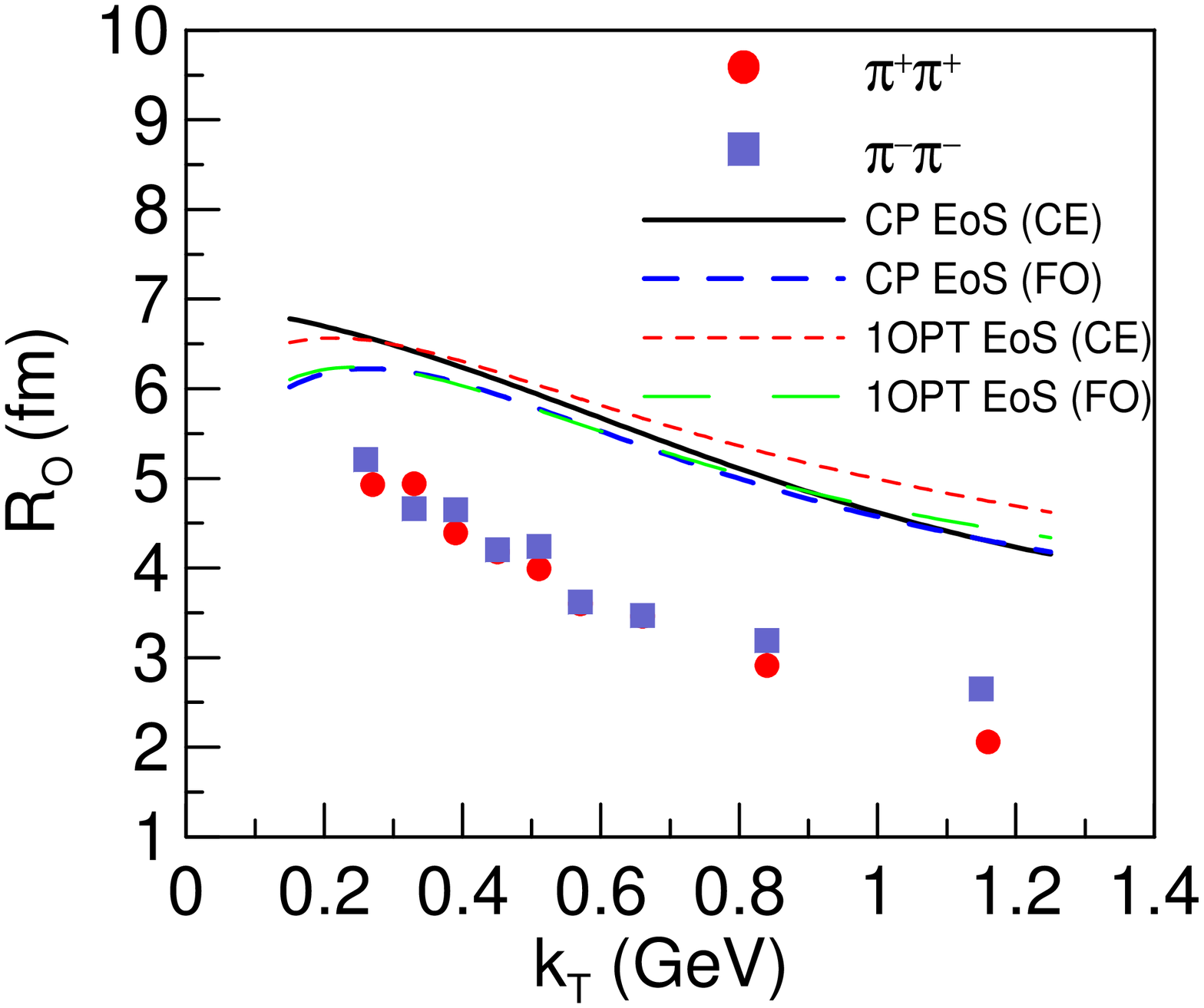}
\caption{$k_T$ depenence of HBT radius $R_o$ for pions in 
 the most central Au+Au at 200A GeV, computed with 
 event-by-event fluctuating IC. The data are from PHENIX 
 Collab.\cite{phenix}.}
\label{fig:Ro}
\end{minipage}
\hspace{\fill}
\begin{minipage}[h]{75mm}
\vspace*{.cm}
\includegraphics*[scale=0.38]{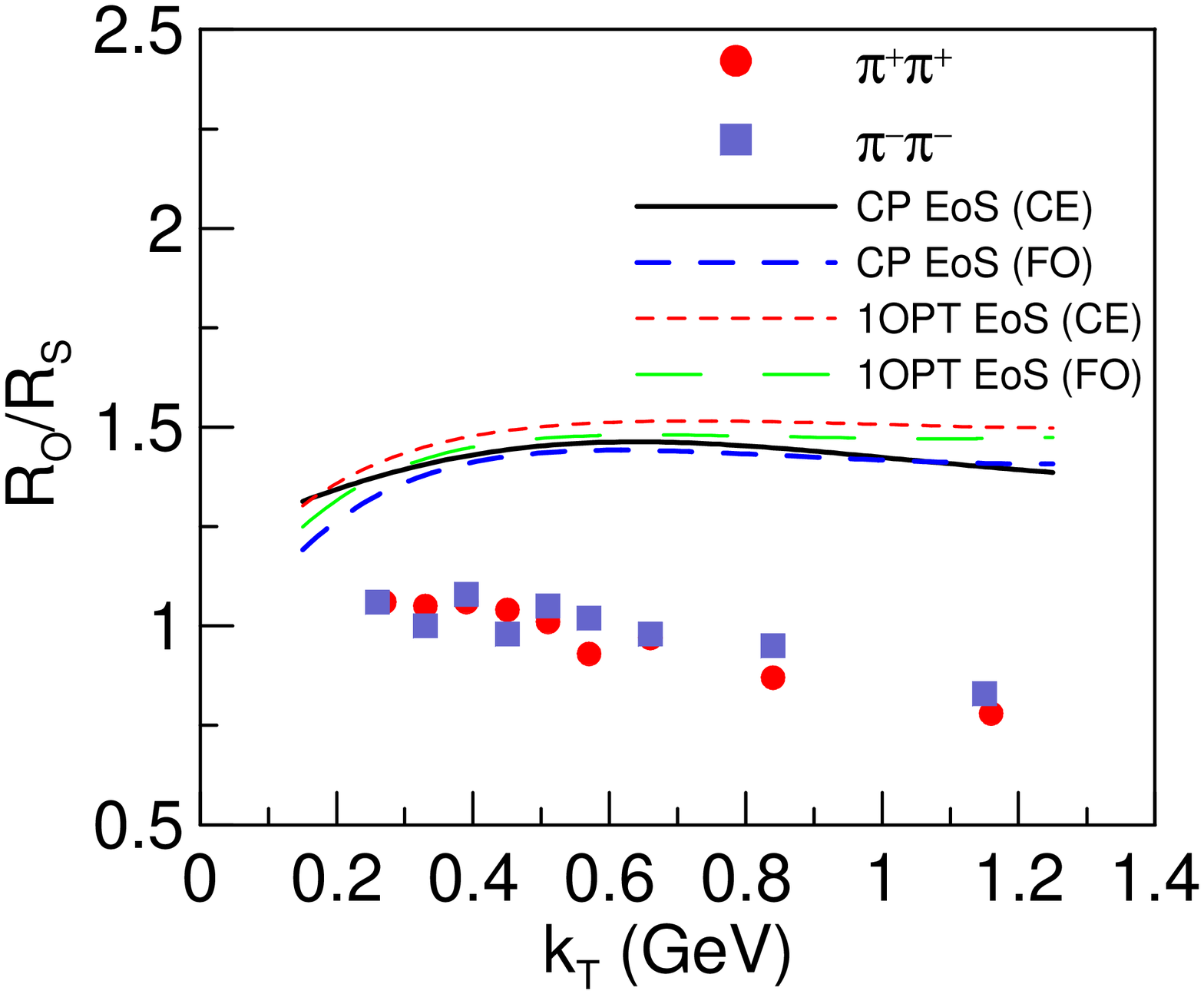}
\caption{$k_T$ depenence of the ratio $R_o/R_s$ for pions in 
 the most central Au+Au at 200A GeV, computed with 
 event-by-event fluctuating IC. The data are from PHENIX 
 Collab.\cite{phenix}.}
\label{fig:RoRs}
\end{minipage}
\end{figure}

Here, we show our results for the HBT radii, in Gaussian 
approximation as used in experimental data analyses, for the 
most central Au+Au collisions at 200A GeV. As seen in 
Figures \ref{fig:RL}, \ref{fig:Rs} and \ref{fig:Ro}, the 
differences between CP EoS results and those for 
1OPT EoS\hfilneg\ 
 
\noindent are small. This was expected for the case of $R_L\,$. 
For $R_s\,$, and especially for $R_o\,$, one sees that CP EoS 
combined with continuous emission gives steeper $k_T$ 
dependence, closer to the data. 
However, there is still numerical discrepancy in this case. 

\section{CONCLUSIONS AND OUTLOOKS}
\label{conclusions}  

In this work, we have introduced a phenomenological 
parametrization of lattice-QCD-inspired equations of state, 
which presents a first-order phase transition at large 
baryonic chemical potential and a crossover behavior at smaller 
chemical potential. By using the initial conditions generated 
by NeXuS event simulator and SPheRIO code for solving the 
hydrodynamic equations, some observables were computed and 
studied the effects of such EoS.
Some of the conclusions are: 
\begin{enumerate}
\item The multiplicity becomes larger for these equations of 
      state in the mid-rapidity;  
\item The $p_T$ distribution becomes flatter. However, the 
      differece is small;  
\item Larger $v_2\,$. {\it Continuous Emission} makes the 
      $\eta$ distribution narrower;  
\item HBT radii slightly closer to data. 
\end{enumerate}
   
In our calculations, the effect of the continuous emission on 
the interacting component has not been taken into account.  A more realistic treatment of this effect probably makes $R_o$ 
smaller, because the duration for particle emission becomes 
smaller in this case. Another improvement we should make is 
the approximations we used for Eq.(\ref{eq:prob1}). 
Cascade treatment of this part is probably a better 
alternative. 

\bigskip

We acknowledge financial support by FAPESP (04/10619-9, 
04/15560-2, 04/13309-0), CAPES/PROBRAL, CNPq, FAPERJ and PRONEX.

\end{document}